\documentstyle[aps,epsfig]{revtex}

\begin{document}
\twocolumn
\title{Kinetic equation for tachyons
\vspace*{0.4em}}
\author{Fuad M. Saradzhev\thanks{email: fuad\_saradzhev@hotmail.com}}
\address{Institute of Physics, National Academy of Sciences of Azerbaijan,
H. Javid pr. 33, 370143 Baku, Azerbaijan\\[0.6\baselineskip]
\parbox{140mm}{\rm \hspace*{1.0em}
{\rm
The tachyonic regime of the quantum fluctuations of a self-interacting
scalar field around its vacuum mean value is studied within a kinetic
approach. We derive a quantum kinetic equation which determines the
time evolution of the momentum distribution function of produced tachyonic
modes and includes memory effects. The back-reaction of the quantum
fluctuations on the vacuum mean field is taken into account, while
their interaction is neglected. We show that the tachyonic modes
do not correspond to real particles and contribute to the decay rate 
of the metastable vacuum state .
\\[0.4\baselineskip]
Pacs Numbers: 11.10-z, 05.20.Dd, 11.10.Ef, 98.80.Cq}
}}

\maketitle

\section{INTRODUCTION}

Quantum fluctuations of a scalar field can enter a tachyonic
regime where their frequency becomes imaginary.  Such regime can occur
if the scalar field is either coupled to a strong stationary
external potential \cite{schiff,tanaka,bert,full} 
or strongly self-interacting with
a potential exhibiting spontaneous symmetry breaking 
\cite{ann,SW,dj}. In the tachyonic regime the system is essentially
restructured. Its effective action develops an imaginary part
\cite{erick}, the fluctuations Hamiltonian becomes unbounded 
from below, while the Hilbert space of states acquires an indefinite
metric \cite{bert}. All these changes are indicative of a new,
metastable phase. 

In the present work, we aim to derive a quantum kinetic equation
describing the production of the tachyonic modes for a
self-interacting neutral massive scalar field. Particle production
in the tachyonic regime has been extensively studied so far in various 
models of spontaneous symmetry breaking \cite{linde}. In these studies,
the occupation number of produced particles has been estimated at the
end of the metastable phase . Herein  we suggest
to study the full time evolution  of the momentum distribution of
the tachyonic modes using a kinetic description.

The decay of the metastable vacuum state has been discussed in
different ways, including the semiclassical approach \cite{cole},
the classical lattice field theory \cite{polo}, the two-particle
irreducible effective action formalism \cite{yoav}. Our approach
is based on the canonical quantization of the tachyonic modes.
 
The plan of the paper is as follows. In Sec.\ref{tachyon} we introduce
the model and identify the tachyonic regime. In Sec.\ref{quant}
we perform the quantization of the tachyonic modes. A quantum kinetic
equation is derived in Sec.\ref{kinetic}. We conclude with summary
in Sec.\ref{summ}.

\section{TACHYONIC REGIME}

\label{tachyon}

We consider a general scalar field model with the Lagrangian
density
\begin{equation}
{\cal L}= \frac{1}{2} ({\partial}_{\mu} {\varphi}) ({\partial}^{\mu}
{\varphi}) - \frac{1}{2} m^2 {\varphi}^2 - V({\varphi}),
\label{1}
\end{equation}
where $V({\varphi})$ is a self-interaction potential which contains
orders ${\varphi}^3$ and higher without derivative terms 
and $m$ is the mass of the scalar field.
The model is defined in a finite volume $L^3$, $-L/2 \leq x_i
\leq L/2$, $i=1,2,3$. The continuum limit is $\frac {1}{L^3}
\sum_{\vec k} \Longrightarrow \int \frac{d^3\vec{k}}{(2{\pi})^3}$.
 
From (\ref{1}) we obtain the Klein-Gordon type equation of motion
for the field ${\varphi}({\vec x},t)$:
\begin{equation}\label{2}
(\Box + m^2) {\varphi} = J \equiv 
- \frac{{\delta}V}{{\delta}{\varphi}},
\end{equation}
where the non-linear current $J$ is also determined by the 
self-interaction.

Following the mean-field approximation, we decompose
${\varphi}(\vec{x},t)$ into its space-homogeneous vacuum mean value 
${\phi}(t)=\langle {\varphi}(\vec{x},t) \rangle$
and fluctuations ${\chi}$ 
\begin{equation}
{\varphi}(\vec{x},t) = {\phi}(t) + {\chi}(\vec{x},t)
\label{3}
\end{equation}
with $\langle {\chi}(\vec{x},t) \rangle =0$.
The mean field is treated as a classical background
field defined with respect to the in-vacuum
$|0 \rangle$ as 
\begin{equation}
{\phi}(t) \equiv \langle {\varphi}(\vec{x},t) \rangle \equiv
\frac{1}{L^3} \int d^3x \langle 0| {\varphi}(\vec{x},t) |0 \rangle,
\label{4}
\end{equation}
so that in the limit $t \to -\infty$  ${\phi}(t) \to 0$, while
the fluctuations are quantized and
take place at all times.

Using Eq.(\ref{3}) provides the following decomposition for the
current
\begin{equation}
J({\phi} + {\chi}) = J({\phi}) + \frac{{\delta}J({\phi})}
{{\delta}{\phi}} {\chi} + \overline{J}({\phi},{\chi}),
\label{5}
\end{equation}
where $\overline{J}({\phi},{\chi})$ includes terms of second and 
higher orders in ${\chi}$,
\begin{equation}
\overline{J}({\phi},{\chi}) = \frac{1}{2}
\frac{{\delta}^2 J({\phi})}{{\delta}{\phi}^2} {\chi}^2 + ...   .
\label{6}
\end{equation}
Substituting Eq.(\ref{3}) also into Eq.(\ref{2}) and
taking the mean value $\langle
... \rangle$  yields the vacuum mean field equation
\begin{equation}
\label{7}
\ddot{\phi} +  m^2 {\phi} - J({\phi})
=\langle \overline{J} \rangle ,
\end{equation}
where the overdot indicates the derivative with respect to time,
while the equation of motion for the quantum fluctuations reads 
\begin{equation}
(\Box + m_{eff}^2) {\chi} = \overline{J} - 
\langle \overline{J} \rangle
\label{8}
\end{equation}
with
\begin{equation}
m_{eff}^2 \equiv m^2 + \frac{{\delta}^2 V({\phi})}{{\delta} {\phi}^2}.
\label{9}
\end{equation}
For $\frac{{\delta}^2 V}{{\delta}{\phi}^2} > 0$, the effective
mass squared is positive at all times. However, if 
$\frac{{\delta}^2 V}{{\delta} {\phi}^2} < 0$, $m_{eff}^2$
becomes negative for $|\frac{{\delta}^2 V}{{\delta} {\phi}^2}|
> m^2$ indicating a tachyonic regime.

In terms of the Fourier components ${\chi}(\vec{k},t)$,
Eq.(\ref{8}) takes the form
\begin{equation}
\ddot{\chi}(\vec{k},t) + {\omega}_k^2(t) {\chi}(\vec{k},t)=
F_{\chi}(\vec{k},t),
\label{10}
\end{equation}
where 
\begin{equation}
F_{\chi}(\vec{k},t) \equiv \overline{J}(\vec{k},t) -
\sqrt{V} \langle \overline{J} \rangle {\delta}_{{\vec k},0}
\label{11}
\end{equation}
and $\overline{J}(\vec{k},t)$ is the Fourier transform of the
current $\overline{J}(\vec{x},t)$,
\begin{equation}
\overline{J}(\vec{k},t) \equiv \frac{1}{L^{3/2}} \int d^3x
e^{-i{\vec k}{\vec x}} \overline{J}(\vec{x},t),
\label{12}
\end{equation}
while
\begin{equation}
{\omega}_k^2(t) \equiv {\vec k}^{~2} + m_{eff}^2(t)
\label{13}
\end{equation}
is the time-dependent frequency squared of the
fluctuations . In the tachyonic regime,  
${\omega}_k^2(t)$ can be negative. 

Whether the system
evolves in the tachyonic or non-tachyonic regime is dynamically
fixed by the time-dependent critical momentum:  
\begin{eqnarray}
\label{14}
{\vec k}_c^{~2} & = &\left\{
\begin{array}{lcl}
\Big| \frac{{\delta}^2 V}{{\delta}{\phi}^2} \Big|
- m^2\,, && \frac{{\delta}^2 V}{{\delta}{\phi}^2}<-m^2\\
0\,, && {\rm otherwise}
\end{array}\right.
\end{eqnarray}
All momentum modes below ${\vec k}_c^{~2}$ are tachyonic.
For $\frac{{\delta}^2 V}{{\delta} {\phi}^2} > - m^2$,
the critical momentum is zero, since the frequency is always
positive and no tachyonic modes can appear.

The system can enter the tachyonic regime in different ways:
gradually when the critical momentum becomes nonzero very
slowly, or discontinuously when tachyonic modes appear
suddenly \cite{anselm,rand,DB2002} on a short time scale. In any case, 
the critical momentum changes in tune with the time dependence
of the vacuum mean field ${\phi}$. If ${\phi}$ oscillates, then
the same momentum mode can change its nature during the time evolution.
 
Eqs. (\ref{7}) and (\ref{10}) are exact, self-consistently coupled
and include back-reactions.
The vacuum mean field modifies the equation for fluctuations
via a time dependent frequency, while the fluctuations 
react back on the vacuum mean field via the source term
$\langle \overline{J} \rangle$ in Eq.(\ref{7}) and on the
fluctuations themselves via the ``external force'' term
$F_{\chi}(\vec{k},t)$ in Eq.(\ref{10}).

\section{QUANTIZATION}

\label{quant}

The Hamiltonian density corresponding to (\ref{1}) is
\begin{equation}
{\cal H} = \frac{1}{2} {\pi}^2 + \frac{1}{2} (\vec{\nabla} {\varphi})^2
+ \frac{1}{2} m^2 {\varphi}^2 + V({\varphi}),
\label{15}
\end{equation}
where ${\pi}$ is the momentum canonically conjugate to ${\varphi}$.
With the decomposition for the potential
\begin{eqnarray}\nonumber
V({\phi}+{\chi}) 
& = & V({\phi}) - J({\phi}){\chi}
+ \frac{1}{2} (m_{eff}^2 - m^2) {\chi}^2\\ 
& + & \overline{V}({\phi},{\chi}),
\label{16}
\end{eqnarray}
orders ${\chi}^3$ and higher being included into $\overline{V}({\phi},
{\chi})$, we deduce from (\ref{15}) the Hamiltonian density governing
the dynamics of the fluctuations
\begin{equation}
{\cal H}_{\chi} \equiv \frac{1}{2} {\pi}_{\chi}^2 +
\frac{1}{2} ( \vec{\nabla} {\chi} )^2 + \frac{1}{2} m_{eff}^2 {\chi}^2
+ \overline{V}({\phi},{\chi}).
\label{17}
\end{equation}

In terms of the Fourier components ${\chi}(\vec{k},t)$ and
${\pi}_{\chi}(\vec{k},t)$, the fluctuations Hamiltonian reads
\begin{eqnarray}\nonumber
&& H_{\chi} = 
\int d^3x {\cal H}_{\chi}\\\nonumber
&& =  \frac{1}{2} \sum_{{\vec k}^{~2} > {\vec k}_c^{~2}}
\Big( {\pi}_{\chi}^{\dagger}(\vec{k},t) {\pi}_{\chi}(\vec{k},t)
+ {\omega}_k^2(t) {\chi}^{\dagger}(\vec{k},t) 
{\chi}(\vec{k},t) \Big)\\\nonumber
&& +  \frac{1}{2} \sum_{{\vec k}^{~2} < {\vec k}_c^{~2}}
\Big( {\pi}_{\chi}^{\dagger}(\vec{k},t) {\pi}_{\chi}(\vec{k},t)
- {\nu}_k^2(t) {\chi}^{\dagger}(\vec{k},t) 
{\chi}(\vec{k},t) \Big)\\
&& +  L^{3/2} \overline{V}(\vec{k}=0,t), 
\label{18}
\end{eqnarray}
where ${\nu}_k^2 \equiv - {\omega}_k^2 > 0$ for
${\vec k}^{~2} < {\vec k}_c^{~2}$, and
\begin{eqnarray}
{\chi}^{\dagger}(\vec{k},t) & = & {\chi}(-\vec{k},t), \\
{\pi}_{\chi}^{\dagger}(\vec{k},t) & = & {\pi}_{\chi}(-\vec{k},t)
\label{19-20}
\end{eqnarray}
for all momentum modes.

The non-tachyonic and tachyonic modes contribution to the
Hamiltonian (\ref{18}) represents a collection of positive
and inverted (repulsive) oscillators, respectively. Both
types of modes are coupled. Their interaction is described
by the last term in Eq.(\ref{18}), $\overline{V}(\vec{k},t)$
being the Fourier transform of the potential $\overline{V}
(\vec{x},t)$.

For the standard, non-tachyonic modes, we introduce the
annihilation and creation operators by
\begin{equation}
{\chi}(\vec{k},t) = {\Gamma}_{\vec{k}}(t) a(\vec{k},t) +
{\Gamma}_{\vec{k}}^{\star}(t) a^{\dagger}(-\vec{k},t),
\label{21}
\end{equation}
and
\begin{equation}
{\pi}_{\chi}(\vec{k},t) = -i{\omega}_k(t) \Big[ 
{\Gamma}_{\vec k}(t) a(-\vec{k},t) -
{\Gamma}_{\vec k}^{\star}(t) a^{\dagger}(\vec{k},t) \Big],
\label{22}
\end{equation}
where
\begin{equation}
{\Gamma}_{\vec{k}}(t) = \frac{1}{\sqrt{2{\omega}_k(t)}}
\exp\{ -i{\Theta}_k({\omega}_k,t) \},
\label{23}
\end{equation}
and ${\Theta}_k({\omega}_k,t)$ is a phase which in the in-limit
takes the form ${\omega}_k^0 t \equiv \sqrt{{\vec k}^2 + m^2} t$.

Eqs.(\ref{21}) and (\ref{22}) are well-known expressions for the
real frequency oscillations. The first term in the Hamiltonian
(\ref{18}) -- we denote it by $H_{\chi}^{nt}$  -- becomes up to a
c-number:
\begin{equation}
H_{\chi}^{nt} =
\sum_{{\vec k}^{~2} > {\vec k}_c^{~2}}
{\omega}_k(t) N^{nt}(\vec{k},t),
\label{24}
\end{equation}
where $N^{nt}(\vec{k},t) \equiv a^{\dagger}(\vec{k},t) a(\vec{k},t)$
is the non-tachyonic modes number density operator.

For the modes with ${\vec k}^{~2} < {\vec k}_c^{~2}$,
${\omega}_k = \pm i{\nu}_k = \pm i \sqrt{{\vec k}_c^{~2} - {\vec k}^2}$
and one of the phase factors in the ansatz (\ref{21}),
${\Gamma}_{\vec k}(t)$ or ${\Gamma}_{\vec k}^{\star}(t)$, grows
exponentially in time. Instead of oscillations we have an exponential
growth of long wavelength quantum fluctuations with momenta
${\vec k}^{~2} < {\vec k}_c^{~2}$. This is the so-called tachyonic
instability \cite{linde,anselm,rand,DB2002}.

Making the transition ${\omega}_k \to {\nu}_k$ in Eq.({22}) yields
the following ansatz for the negative frequency squared fluctuations
\begin{eqnarray}\nonumber
&& {\chi}({\vec k},t) \to {\chi}_t({\vec k},t) \\
&& = \frac{1}{\sqrt{2{\nu}_k}} \Big( e^{{\vartheta}_k} a({\vec k},t)
+ e^{-{\vartheta}_k} a^{\dagger}(-{\vec k},t) \Big),
\label{25}
\end{eqnarray}
where ${\vartheta}_k({\nu}_k,t) = -i{\Theta}_k({\omega}_k,t)$.
Introducing
\begin{eqnarray}
{\sigma}_1({\vec k},t) & \equiv & \frac{1}{\sqrt{2{\nu}_k}}
\cosh{\vartheta}_k \cdot \Big( a({\vec k},t) + a^{\dagger}(-{\vec k},t)
\Big),\\
{\sigma}_2({\vec k},t) & \equiv & - \frac{1}{\sqrt{2{\nu}_k}}
\sinh{\vartheta}_k \cdot \Big( a({\vec k},t) - a^{\dagger}(-{\vec k},t)
\Big),
\label{26-27}
\end{eqnarray}
which obey the hermiticity condition ${\sigma}_{1(2)}^{\dagger}
({\vec k},t) = {\sigma}_{1(2)}(-{\vec k},t)$, we rewrite Eq.(\ref{25})
as
\begin{equation}
{\chi}_t({\vec k},t) = {\sigma}_1({\vec k},t) + i {\sigma}_2({\vec k},t)
\label{28}
\end{equation}
with ${\chi}_t^{\dagger}({\vec k},t) \neq {\chi}_t(-{\vec k},t)$,
i.e. the ansatz (\ref{25}) is non-Hermitian.

The canonically conjugate momentum is transformed as
\begin{equation}
{\pi}_{\chi}({\vec k},t) \to {\pi}_{{\chi},t}({\vec k},t) =
{\pi}_{{\sigma}1}({\vec k},t) + i{\pi}_{{\sigma}2}({\vec k},t),
\label{29}
\end{equation}
where
\begin{eqnarray}
{\pi}_{{\sigma}1}(\vec{k},t) & \equiv &
{\nu}_k \coth{\vartheta}_k \cdot {\sigma}_2^{\dagger}(\vec{k},t),\\
{\pi}_{{\sigma}2}(\vec{k},t) & \equiv &
- {\nu}_k \tanh{\vartheta}_k \cdot {\sigma}_1^{\dagger}(\vec{k},t).
\label{30-31}
\end{eqnarray}
The commutation relations for ${\sigma}_1$,${\sigma}_2$-fields are
\begin{equation}
{\Big[ {\sigma}_1(\vec{k},t) , {\sigma}_2({\vec p},t) \Big]}_{-}
= \frac{i}{2{\nu}_k} \sinh2{\vartheta}_k \cdot {\delta}_{{\vec k},
-{\vec p}},
\label{32}
\end{equation}
all other commutators vanishing.

Analytically continuing the ansatz (\ref{21}) in the frequency
to imaginary values leads therefore to a non-Hermitian field.
This is not acceptable because we require the hermiticity
conditions, Eqs.(19)-(20), to be valid for all momentum
modes and at all steps of our consideration. In addition, such
non-Hermitian field is known to violate causality \cite{sudar}.

To define the Hermitian tachyonic fluctuations we can use
either ${\sigma}_1({\vec k},t)$ or ${\sigma}_2({\vec k},t)$
instead of ${\chi}_t({\vec k},t)$. Without loss of generality,
we choose ${\sigma}_1({\vec k},t)$ and introduce the field
\begin{equation}
{\sigma}_t({\vec k},t) \equiv \frac{1}{\cosh{\vartheta}_k}
{\sigma}_1({\vec k},t).
\label{33}
\end{equation}
Its canonically conjugate momentum is
\begin{equation}
{\pi}_{{\sigma},t}({\vec k},t) \equiv \frac{1}{\cosh{\vartheta}_k}
{\pi}_{{\sigma},1}({\vec k},t).
\label{34}
\end{equation}
With Eqs.(\ref{33}) and (\ref{34}), the second term in the
Hamiltonian (\ref{18}) takes the form
\begin{equation}
H_{\chi}^t = \sum_{{\vec k}^{~2} < {\vec k}_c^{~2}}
{\nu}_k(t) N^t({\vec k},t),
\label{35}
\end{equation}
where
\begin{equation}
N^t({\vec k},t) \equiv - \frac{1}{2} \Big( a^{\dagger}({\vec k},t)
a^{\dagger}(-{\vec k},t) + a(-{\vec k},t) a({\vec k},t) \Big).
\label{36}
\end{equation}

Since the spectrum of an inverted oscillator is purely continuous,
the tachyonic modes are not really ``particle'' ones \cite{guth}. 
In contrast
with the case of the standard, non-tachyonic modes where the
eigenfunctions of $H_{\chi}^{nt}$ coincide with those of the
number operator, the tachyonic modes are not eigenoperators of
\begin{equation}
N^t \equiv \sum_{{\vec k}^{~2}<{\vec k}_c^{~2}}
N^t({\vec k},t),
\label{37}
\end{equation}
namely
\begin{eqnarray}
{\Big[ N^t , a({\vec k},t) \Big]}_{-} & = & a^{\dagger}
(-{\vec k},t), \\
{\Big[ N^t , a^{\dagger}({\vec k},t) \Big]}_{-} & = &
- a(-{\vec k},t),
\label{38-39}
\end{eqnarray}
so that $a({\vec k},t)$,$a^{\dagger}({\vec k},t)$ in Eq.(\ref{35})
can not be viewed as creation and annihilation operators.

However, once complex values are allowed for energy, the particle 
interpretation can be kept for the tachyonic
modes as well~.
Let us introduce
\begin{equation}
{\alpha}_{1(2)}({\vec k},t) \equiv 
\frac{1 \mp i}{2} a^{\dagger}(-{\vec k},t)
+ \frac{1 \pm i}{2} a({\vec k},t), 
\label{40}
\end{equation}
where the upper signs correspond to the subscript 1 and the lower ones
to 2.
These new mode operators are Hermitian and fulfill the algebra
\begin{eqnarray}
&& {\Big[ {\alpha}_1({\vec k},t) , {\alpha}_1^{\dagger}({\vec p},t) \Big]}_{-}
= {\Big[ {\alpha}_2({\vec k},t) , {\alpha}_2^{\dagger}({\vec k},t) \Big]}_{-}
=0, \\
&& {\Big[ {\alpha}_1({\vec k},t) , {\alpha}_2^{\dagger}({\vec p},t) \Big]}_{-}
= i {\delta}_{{\vec k},{\vec p}}.
\label{41-42}
\end{eqnarray}
The Fock representation for the algebra (41)-(42) is constructed
by using an indefinite metric . Indeed, if $|0;t \rangle$
is an instantaneous vacuum state defined by
\begin{equation}
{\alpha}_1({\vec k},t) |0;t \rangle =
{\alpha}_2({\vec k},t) |0;t \rangle =0
\hspace{5 mm}
{\rm for}
\hspace{5 mm}
(k_i)>0,
\label{43}
\end{equation}
where $(k_i)=(k_1,k_2,k_3)$, then for the excited states
\begin{equation}
|{\alpha}_{1(2)};{\vec k},t \rangle \equiv
{\alpha}_{1(2)}^{\dagger}({\vec k},t) |0;t \rangle,
\hspace{5 mm}
(k_i)>0,
\label{44}
\end{equation}
the inner product is vanishing or imaginary,
\begin{eqnarray}
\langle {\alpha}_{1(2)};{\vec k},t | {\alpha}_{1(2)};{\vec p},t \rangle
& = & 0,\\
\langle {\alpha}_1;{\vec k},t | {\alpha}_2;{\vec p},t \rangle
& = & i {\delta}_{{\vec k},{\vec p}}.
\label{45-46}
\end{eqnarray}
The indefinite inner product is related to the existence of
associated eigenvectors of $H_{\chi}$ \cite{keld}.

The density of $N^t$ becomes
\begin{eqnarray}\nonumber
N^t({\vec k},t) & = & 
- i N_{\alpha}^t({\vec k},t)\\
& \equiv &
- \frac{1}{2} \Big( {\alpha}_1^{\dagger}({\vec k},t)
{\alpha}_2({\vec k},t) + {\alpha}_2^{\dagger}({\vec k},t) 
{\alpha}_1({\vec k},t)
\Big),
\label{47}
\end{eqnarray}
${\alpha}_{1(2)}({\vec k},t)$ being eigenoperators of
\begin{equation}
N_{\alpha}^t \equiv \sum_{\stackrel{{\vec k}^{~2}<{\vec k}_c^{~2}}
{(k_i)>0}} 2N_{\alpha}^t({\vec k},t)
\label{48}
\end{equation}
with real eigenvalues,
\begin{equation}
{\Big[ N_{\alpha}^t , {\alpha}_{1(2)}({\vec k},t) \Big]}_{-} = 
\mp  {\alpha}_{1(2)}({\vec k},t) .
\label{49}
\end{equation}
For the instantaneous vacuum, $N_{\alpha}^t |0;t \rangle=0$,
while for the excited states $N_{\alpha}^t$ counts excitations.
For the state
\begin{eqnarray}\nonumber
&& |n{\alpha}_2; {\vec k}_1, {\vec k}_2, ...,{\vec k}_n;t \rangle\\\nonumber
&& \equiv {\alpha}_2^{\dagger}({\vec k}_1,t) 
{\alpha}_2^{\dagger}({\vec k}_2,t) \cdot
... \cdot {\alpha}_2^{\dagger}({\vec k}_n,t) |0;t \rangle,\\
&& {\rm all}
\hspace{5 mm}
(k_{1,i},...,k_{n,i})>0,
\label{50}
\end{eqnarray}
for instance,
\begin{equation}
N_{\alpha}^t |n{\alpha}_2;{\vec k}_1, {\vec k}_2,...,
{\vec k}_n ;t \rangle
= n |n{\alpha}_2;{\vec k}_1, {\vec k}_2, ...,{\vec k}_n ;t \rangle,
\label{51}
\end{equation}
i.e. $N_{\alpha}^t$ plays the role of the ``number operator''.

In the space with indefinite metric, the Hamiltonian  $H_{\chi}^t$
is pseudoadjoint \cite{bert} and its eigenvalues 
are imaginary. If $| \varepsilon ;t \rangle$ is an 
eigenstate of $H_{\chi}^t$ with eigenvalue $\varepsilon$,
then for the state ${\alpha}_2^{\dagger}({\vec k},t)| \varepsilon ;t \rangle$
we obtain
\begin{equation}
H_{\chi}^t {\alpha}_2^{\dagger}({\vec k},t) | \varepsilon ;t \rangle =
\Big( {\varepsilon} + i{\nu}_k \Big) {\alpha}_2^{\dagger}({\vec k},t)
| \varepsilon ;t \rangle,
\label{52}
\end{equation}
i.e. ${\alpha}_2^{\dagger}({\vec k},t) | \varepsilon ;t \rangle$ is also
an eigenstate of $H_{\chi}^t$ with the eigenvalue shifted by
$i{\nu}_k$.
Neglecting for a moment the third term in the right-hand sise
of Eq.(\ref{18}), we see
that the eigenvalues of the total Hamiltonian $H_{\chi}^{nt} +
H_{\chi}^t$ are complex, the corresponding eigenfunctions
representing unstable states.

\section{KINETIC EQUATION}

\label{kinetic}

In the mean-field approximation, the quantum fluctuations
are treated perturbatively. This is necessary, in particular,
for the derivation of the kinetic equation. One of basic points
of the kinetic formulation is the postulate of asymptotic
completeness \cite{SR1980}. The postulate specifies the set 
of possible states
of the system in the infinite past as a complete set of states
of freely moving non-interacting particles. For the system with
a self-interaction, this postulate can be applied only in a few 
lower orders of perturbations when the interaction potential
vanishes in the in-limit due to the vanishing of the vacuum mean
field~. In higher orders, the quantum
fluctuations dominate, the corresponding terms in the interaction
potential surviving in the infinite past.

We limit our consideration to the third order term in
$\overline{V}({\phi},{\chi})$ neglecting all higher orders.
In addition, we use a Hartree-type approximation that in the
second and third orders consists of the factorization
\begin{equation} 
{\chi}^2 \to \langle {\chi}^2 \rangle,
\hspace{5 mm}
{\chi}^3 \to 3 \langle {\chi}^2 \rangle {\chi}.
\label{53}
\end{equation}

For the non-tachyonic modes, the form of the kinetic equation
is well-known and was given in different models \cite{klug,DB2002}.
It determines the time evolution of the 
occupation number density
\begin{equation}
{\cal N}^{nt}({\vec k},t) \equiv \langle 0| N^{nt}({\vec k},t)
|0 \rangle
\label{54}
\end{equation}
which defines the number of particles of a given state characterized
by the momentum ${\vec k}^{~2}>{\vec k}_c^{~2}$ at time $t$.
An increase in the occupation number density
is interpreted as particle production. 

Herein we focus on the time evolution of
\begin{equation}
{\cal N}^t({\vec k},t) \equiv \langle 0| N_{\alpha}^t({\vec k},t)
|0 \rangle
\label{55}
\end{equation}
which defines the momentum distribution of the tachyonic modes.
We start with the tachyonic Hamiltonian equations of motion
\begin{eqnarray}
\dot{\sigma}_t({\vec k},t) & = &
{\pi}_{{\sigma},t}^{\dagger}({\vec k},t),\\
\dot{\pi}_{{\sigma},t}({\vec k},t) & = &
{\nu}_k^2 {\sigma}^{\dagger}_t({\vec k},t) +
\overline{J}(-{\vec k},t),
\label{56-57}
\end{eqnarray}
where the current $\overline{J}(-\vec{k},t)$ represents
the self-interaction potential contribution. With the factorization
(\ref{53}), the self-interaction potential and current take the form
\begin{equation}
\overline{V}({\vec k}=0,t) = -\frac{1}{2}
\frac{{\delta}^2J({\phi})}{{\delta}{\phi}^2}
\langle {\chi}^2 \rangle {\sigma}_t(0,t)
\label{58}
\end{equation}
and
\begin{equation}
\overline{J}(\vec{k},t) = \frac{1}{2} L^{3/2}
\frac{{\delta}^2 J({\phi})}{{\delta} {\phi}^2}
\langle {\chi}^2 \rangle {\delta}_{{\vec k},0},
\label{59}
\end{equation}
respectively. Using the relations
\begin{equation}
{\alpha}_{1(2)}({\vec k},t) =  \sqrt{\frac{{\nu}_k}{2}}
\Big( {\sigma}_t({\vec k},t) \mp \frac{1}{{\nu}_k}
{\pi}_{{\sigma},t}^{\dagger}({\vec k},t) \Big) 
\label{60}
\end{equation}
yields then the equations for ${\alpha}_{1(2)}({\vec k},t)$:
\begin{eqnarray}\nonumber
\dot{\alpha}_{1(2)}({\vec k},t) & \pm & {\nu}_k {\alpha}_{1(2)}({\vec k},t) - 
\frac{\dot{\nu}_k}{2{\nu}_k} {\alpha}_{2(1)}({\vec k},t)\\
& = & \mp \frac{1}{\sqrt{2{\nu}_k}} \overline{J}({\vec k},t).
\label{61}
\end{eqnarray}

Taking next the time derivative of ${\cal N}^t({\vec k},t)$
we find:
\begin{equation}
\dot{\cal N}^t({\vec k},t) = \frac{\dot{\nu}_k}{2{\nu}_k}
\Big( C_{1}({\vec k},t) + C_{2}({\vec k},t) \Big),
\label{62}
\end{equation}
where we have defined the time-dependent one-particle correlation
functions
\begin{equation}
C_{1(2)}({\vec k},t) \equiv 
\langle 0| {\alpha}_{1(2)}
(-{\vec k},t) {\alpha}_{1(2)}({\vec k},t) |0 \rangle .
\label{63}
\end{equation}
Since $\langle {\chi}({\vec x},t) \rangle =0$, the vacuum
expectation values for the zero momentum mode operators
${\sigma}_t(0,t)$ and ${\pi}_{{\sigma},t}(0,t)$ are equal
to zero, and , as a result, the current $\overline{J}({\vec k},t)$
drops out of Eq.(\ref{62}).

The functions $C_{1(2)}({\vec k},t)$ 
obey the equations
\begin{equation}
\dot{C}_{1(2)}({\vec k},t) = \frac{\dot{\nu}_k}{{\nu}_k}
{\cal N}^t({\vec k},t) \mp 2{\nu}_k C_{1(2)}({\vec k},t) .
\label{64}
\end{equation}
Their formal solution is
\begin{eqnarray}\nonumber
C_{1(2)} ({\vec k},t) 
& = & \int_{t_0}^t dt^{\prime} \frac{\dot{\nu}_k(t^{\prime})}
{{\nu}_k(t^{\prime})} {\cal N}^t({\vec k},t^{\prime}) \\
& \times & \exp\{ \pm 2( {\vartheta}_k^{ad}(t^{\prime})
- {\vartheta}_k^{ad}(t) ) \},
\label{65}
\end{eqnarray}
where
\begin{equation}
{\vartheta}_k^{ad}(t) \equiv \int_{t_0}^t 
dt^{\prime} {\nu}_k(t^{\prime})
\label{66}
\end{equation}
and $t_0$ is a moment of time at which the tachyonic regime starts.
If $t_0 = -\infty$ , then the in-vacuum can be chosen as an initial 
state of the system.

Although we have not assumed that the frequency ${\nu}_k$ varies
adiabatically slowly in time and the phase ${\vartheta}_k$ in the
ansatz (\ref{25}) is a general function of ${\omega}_k$ and $t$,
it is just the ``adiabatic'' phase (\ref{66}), i.e. the phase which looks
exactly like the one in the adiabatic case, that enters this solution.
Substituting it  into Eq.(\ref{62}), we obtain a closed
equation for ${\cal N}^t({\vec k},t)$:
\begin{eqnarray}\nonumber
\dot{\cal N}^t(\vec{k},t) & = &
\frac{\dot{\nu}_k}{{\nu}_k}
\int_{t_0}^{t} dt^{\prime}  \frac{\dot{\nu}_k(t')}
{{\nu}_k(t')} {\cal N}^t(\vec{k},t^{\prime})\\  
& \times &
\cosh\Big[ 2{\vartheta}_k^{ad}(t^{\prime}) - 2{\vartheta}_k^{ad}(t) \Big]
\label{67}
\end{eqnarray}
This kinetic equation determines the time evolution of the
momentum distribution of the tachyonic modes produced in the fluctuations
of the scalar field. As seen from the definition (\ref{47}), the
tachyonic modes production is symmetric in the momentum space, 
${\cal N}^t({\vec k},t)={\cal N}^t(-{\vec k},t)$ for all times $t$.

Eq.(\ref{67}) has non-Markovian character due to the explicit
dependence of its right-hand side - the source term~${ -}$ on the
time evolution of ${\cal N}^t({\vec k},t)$ and therefore involves
memory effects starting from $t_0$. For the real particle 
modes, the source term is known
to contain a time integration over the statistical factor
$(1 \pm 2{\cal N}({\vec k},t))$, where the plus sign corresponds to
bosons and the minus one to fermions \cite{klug}.  For the
tachyonic modes , this
factor reduces to $2{\cal N}({\vec k},t)$ reflecting once more the
fact that tachyons are not real particles .

In our approximation, the vacuum mean field equation becomes
\begin{equation}
\ddot{\phi} + m^2{\phi} = J(\phi) + \frac{1}{2}
\frac{{\delta}^2 J}{{\delta} {\phi}^2} \langle {\chi}^2 \rangle,
\label{68}
\end{equation}
where the $\langle {\chi}^2 \rangle$-term  
represents the back-reaction
of the fluctuations on the vacuum mean field and provides damping
of the oscillations of ${\phi}$. The initial conditions for both
Eqs.(\ref{67}) and (\ref{68}) are specified by the model under
study.

The vacuum mean value of ${\chi}^2$ is given by
\begin{eqnarray}\nonumber
\langle {\chi}^2 \rangle 
& = & \frac{1}{L^3} \sum_{{\vec k}^{~2}>{\vec k}_c^{~2}} 
\langle 0| {\chi}(\vec{k},t) {\chi}(-\vec{k},t) |0 \rangle \\
& + & \frac{1}{L^3} \sum_{{\vec k}^{~2}<{\vec k}_c^{~2}}
\langle 0| {\sigma}_t({\vec k},t) {\sigma}_t(-{\vec k},t)
|0 \rangle ,
\label{69}
\end{eqnarray}
the bi-linear operator expressions here being assumed to be
normal-ordered with respect to the instantaneous vacuum  state.
Both types of modes, tachyonic and non-tachyonic, contribute
to $\langle {\chi}^2 \rangle$, so a proper inclusion of the
back-reactions effects can be achieved only by the complete 
treatment of all momentum modes.

Taking the vacuum expectation value of the fluctuations
Hamiltonian yields
\begin{equation}
\langle 0| H_{\chi} |0 \rangle =
E_{\chi} - i \frac{{\Gamma}_{\chi}}{2},
\label{70}
\end{equation}
where the non-tachyonic modes contribute to the energy of
the metastable vacuum state
\begin{equation}
E_{\chi} \equiv \sum_{{\vec k}^{~2}>{\vec k}_c^{~2}}
{\omega}_k(t) {\cal N}^{nt}({\vec k},t),
\label{71}
\end{equation}
while the tachyonic ones to its decay rate,
\begin{equation}
{\Gamma}_{\chi} \equiv 2 \sum_{{\vec k}^{~2}<{\vec k}_c^{~2}}
{\nu}_k(t) {\cal N}^t({\vec k},t).
\label{72}
\end{equation}
%

\section{SUMMARY}

\label{summ}

For the model of a self-interacting scalar field, we have
derived a non-Markovian quantum kinetic equation determining
the momentum distribution of the tachyonic modes. These modes
are produced in quantum fluctuations of the scalar field
around its vacuum mean value when the system is in a metastable
phase. The kinetic and vacuum mean field equations are coupled,
the latter including the back-reaction term, while the
collisions effects are neglected.

Despite the fact that the fluctuations Hamiltonian is not bounded from below
in the tachyonic regime, the conservation of energy prevents
any catastrophic production of tachyons. If the system starts
in a false, metastable vacuum state and then undergoes the transition
to a lower energy density , stable one, the tachyonic regime stops as
soon as all the potential energy of the false vacuum state 
is transferred into the quantum fluctuations. We have shown that the tachyonic
modes contribute to the decay rate  of this state, so
their intensive production results in its rapid decay.

The kinetic equation obtained is hoped to be useful for the
numerical study of the tachyonic modes production in various
problems, in particular, in cosmology \cite{linde} 
and heavy-ion collisions \cite{anselm,rand,DB2002}.
The complete study requires  the inclusion of
higher orders effects when the quantum fluctuations interact
with each other. Its realization within the kinetic formulation
would provide further insight into the dynamics of the tachyonic
regime.

\section*{Acknowledgment}

This research was supported by the
Deutsche Forschungsgemeinschaft under project number 436 ASB 17/1/02.
The author thanks D.B.~Blaschke and  V.G.~Morozov for discussions.
He also acknowledges hospitality at the University of Rostock
where this work was completed.


\begin{thebibliography}{99}
\bibitem{schiff}
L.I.~Schiff, H.~Snyder, and J.~Weinberg,
Phys.\ Rev.\ {\bf 15} (1940) 315.
\bibitem{tanaka}
S.~Tanaka,
Progr.\ Theor.\ Phys.\ {\bf 24} (1960) 171.
\bibitem{bert}
B.~Schroer and J.A.~Swieca,
Phys.\ Rev.\ {\bf D 2} (1970) 2938;
B.~Schroer, 
Phys.\ Rev.\ {\bf D 3} (1971) 1764.
\bibitem{full}
S.A.~Fulling,
Phys.\ Rev.\ {\bf D 14} (1976) 1939.
\bibitem{ann}
D.A.~Kirzhnits and A.D.~Linde,
Phys.\ Lett.\ {\bf B 42} (1972) 471;
Ann.\ Phys.\ (NY) {\bf 101} (1976) 195.
\bibitem{SW}
S.~Weinberg,
Phys.\ Rev.\ {\bf D 9} (1974) 3357.
\bibitem{dj}
L.~Dolan and R.~Jackiw,
Phys.\ Rev.\ {\bf D 9} (1974) 3320.
\bibitem{erick}
E.J.~Weinberg and A.~Wu,
Phys.\ Rev.\ {\bf D 8} (1987) 2474.
\bibitem{linde}
G.~Felder et al.,
Phys.\ Rev.\ Lett.\ {\bf 87} (2001) 011601;
G.~Felder, L.~Kofman, and A.~Linde,
Phys.\ Rev.\ {\bf D 64} (2001) 123517 and references therein.
\bibitem{cole}
S.~Coleman,
Phys.\ Rev.\ {\bf D 15} (1977) 2929;
S.~Coleman and F.~De Luccia,
Phys.\ Rev.\ {\bf D 21} (1980) 3305.
\bibitem{polo}
Sz.~Bors\'anyi et al.,
Phys.\ Rev.\ {\bf D 62} (2000) 085013.
\bibitem{yoav}
Y.~Bergner and L.M.A.~Bettencourt,
hep-ph-0206053.
\bibitem{anselm}
A.A.~Anselm and M.G.~Ryskin,
Phys.\ Lett.\ {\bf B 266} (1991) 482;
K.~Rajagopal and F.~Wilczek,
Nucl.\ Phys.\ {\bf B 399} (1993) 395; {\bf B 404} (1993) 577;
D.~Boyanovsky, H.J.~de Vega, and R.~Holman,
Phys.\ Rev.\ {\bf D 51} (1995) 734.
\bibitem{rand}
J.~Schaffner-Bielich and J.~Randrup,
Phys.\ Rev.\ {\bf C 59} (1999) 3329;
J.~Randrup,
Heavy Ion Phys.\ {\bf 9} (1999) 289;
Phys.\ Rev.\ {\bf C 62} (2000) 064905.
\bibitem{DB2002}
D.B.~Blaschke et al.,
Phys.\ Rev.\ {\bf D 65} (2002) 054039.
\bibitem{sudar}
M.E.~Arons and E.C.G.~Sudarshan,
Phys.\ Rev.\ {\bf 173} (1968) 1622;
J.~Dhar and E.C.G.~Sudarshan,
Phys.\ Rev.\ {\bf 174} (1968) 1808.
\bibitem{guth}
A.H.~Guth and S.-Y. Pi,
Phys.\ Rev.\ {\bf D 32} (1985) 1899;
G.~Barton,
Ann.\ Phys.\ (NY) {\bf 166} (1986) 322.
\bibitem{keld}
M.V.~Keldysh, 
Russian Math.\ Surveys {\bf 26} (4) (1971) 15;\\
G.V.~Radzievski,
ibid {\bf 37} (2) (1982) 81.
\bibitem{SR1980}
S.R.~deGroot, W.A.~van Leeuwen, and C.G.~van Weert, {\it Relativistic
Kinetic Theory} (North-Holland, Amsterdam, 1980).
\bibitem{klug}
Y.~Kluger, E.~Mottola, and J.M.~Eisenberg,
Phys.\ Rev. {\bf D 58} (1998) 125015 and references therein;
S.M.~Schmidt et al.,
Int.\ J.\ Mod.\ Phys.\ {\bf E 7} (1998) 709;
Phys.\ Rev.\ {\bf D 59} (1999) 094005.
\end{thebibliography}
\end{document}